\documentclass[graybox]{svmult}

\usepackage{mathptmx}       
\usepackage{helvet}         
\usepackage{courier}        
\usepackage{type1cm}        
\usepackage{makeidx}         
\usepackage{graphicx}        
\usepackage{multicol}        
\usepackage[bottom]{footmisc}

\usepackage{latexsym,amsmath,amssymb,amsfonts} 
\usepackage{mathrsfs}        
\usepackage{subeqnarray}     

\newcommand{\ud}{\ensuremath{\mathrm{d}}}           

\makeindex             

\begin{document}

\title*{Chaos in the Kepler problem with quadrupole perturbations}
\author{Gabriela Depetri and Alberto Saa}
\institute{Gabriela Depetri \at Instituto de F\'\i sica Gleb Wataghin, Universidade Estadual de Campinas, 
13083-859 Campinas, SP, Brazil, \email{gdepetri@fma.if.usp.br}
\and Alberto Saa \at Departamento de Matem\'atica Aplicada,   Universidade Estadual de Campinas, 
13083-859 Campi\-nas, SP, Brazil, \email{asaa@ime.unicamp.br}}
%
%
\maketitle

\abstract*{We use the Melnikov integral method to prove that the
  Hamiltonian flow on the zero-energy manifold for the Kepler problem
  perturbed by a quadrupole moment is chaotic, irrespective of the
  perturbation being of prolate or oblate type. This result helps to elucidate 
  some  recent conflicting works in the physical literature based on
  numerical simulations.  
} 

\abstract{We use the Melnikov integral method to prove that the
  Hamiltonian flow on the zero-energy manifold for the Kepler problem
  perturbed by a quadrupole moment is chaotic, irrespective of the
  perturbation being of prolate or oblate type. This result helps to elucidate 
  some  recent conflicting works in the physical literature based on
  numerical simulations.  
}

\section{Introduction}
\label{sec:1}

The Kepler two-body problem has been a splendid inspiration for physicists and
mathematicians for the last three centuries (see, for instance, Chapter 9 of \cite{kepler}). Many works, in particular,
have been devoted to the study of the onset of chaos in the perturbed
Kepler problem (see, for a recent review, \cite{JMP} and the
references therein). For astronomical  and astrophysical applications,
it is natural to consider the weak field approximation in which the
gravitational field of a body is decomposed into a multipole expansion. The
original Kepler problem corresponds to the case where only the first
expansion term, the monopole, is present. The next term in the
expansion, the dipole term, is known to give origin to integrable
motion, see Chapter 7 of \cite{dipole} and Section 2 below. The quadrupole term is usually considered as the
simplest perturbation to the Newtonian potential which could lead to
chaotic motion in the Kepler problem (see, for instance,
\cite{GL}). By employing  the usual cylindrical coordinates $(r,z,\phi)$ around
the gravitational center, the simplest quadrupole perturbation to the Newtonian potential reads 
\begin{equation}
\label{pot}
U(r,z) = -\frac{\alpha}{\sqrt{r^2+z^2}} -
\frac{q}{2}\frac{2z^2-r^2}{\left(r^2+z^2\right)^{5/2}}, 
\end{equation}
where $\alpha$ and $q$ stands for, respectively, the monopole
intensity (proportional to the total gravitational mass) and the quadrupole
intensity. The cylindrical coordinates are assumed to be adjusted to the
quadrupole direction. Two qualitative distinct cases can be
distinguished for the potential (\ref{pot}). Oblate deformations, as
those ones of rotating deformed bodies, corresponds to $q<0$, whereas
prolate deformations, as cigar-like mass distributions, to $q>0$. The
study of the integrability of a test body motion under action of the potential
(\ref{pot}) is a long standing problem, with 
substantially 
relevance to astronomy and astrophysics \cite{BP}. 

In \cite{GL}, a numerical study of bounded trajectories is reported suggesting that the motion
under prolate perturbations would be indeed chaotic while, on the other hand,
oblate perturbations would correspond to an integrable case. Such conclusion would be rather
puzzling since it is known that, for disk-like perturbation (which
could be understood as extreme oblate perturbations), bounded oblique 
orbits are known to 
be chaotic \cite{SV,S}. This qualitative differences for the
cases $q>0$ and $q<0$ is attributed in \cite{GL} to some qualitative
differences in the saddle points of the effective potential, but it is
also known that such kind of local argument leads typically to
conditions that are not sufficient neither necessary to the appearance
of chaos in theses systems \cite{AS}. More recently, a new numerical
study suggesting that the oblate perturbations would   also give origin to
bounded chaotic orbits 
has appeared  \cite{LCF}. Here, we explore these conflicting results by
applying the Melnikov integral method \cite{Melnikov} for the
parabolic orbits \cite{revisited} (the zero-energy manifold) of (\ref{pot}). We prove
the quadrupole perturbations effectively give rise to chaotic motion on the zero-energy manifold, irrespective
of the perturbation being of prolate $(q>0)$ or oblate $(q<0)$ type.

\section{The Melnikov Conditions}

The Hamiltonian associated to the motion of a test body of unit mass
under the action of the potential (\ref{pot}) is given by 
\begin{equation}
\label{H}
H = \frac{1}{2}\left(p_r^2 + p_z^2\right) + \frac{L_z^2}{2r^2} + U(r,z),
\end{equation}
where $(r,p_r)$ and $(z,p_z)$ stands for the usual canonical cylindrical 
coordinates and $L_z$ is the (conserved) angular momentum around the
$z$ axis. The Hamiltonian $H$ is itself a conserved quantity and the
integrability of the Hamiltonian flow governed by (\ref{H})
corresponds to the celebrated problem of the existence of the third
isolating conserved integral of motion \cite{BP}. In order the write
the quadrupole perturbation in (\ref{H}) conveniently, let us
introduce the new variables $(R,\theta)$  
\begin{eqnarray}
\left\{
\begin{array}{l}
r = R\cos\theta, \\
z = R\sin\theta,
\end{array}
\right.
\end{eqnarray}
which leads to
\begin{equation}
\label{ham}
H = H_0 + qW_1(R) + qW_2(R,\theta),
\end{equation}
where
\begin{equation}
\label{H_0}
H_0 = \frac{1}{2}\left(p_R^2 + \frac{p_\theta^2}{R^2} \right) +
\frac{L_z^2}{2R^2\cos^2\theta} - \frac{\alpha}{R}, 
\end{equation}
with $(R,p_R)$ and $(\theta,p_\theta)$ standing for the usual
canonical coordinates, and 
\begin{equation}
\label{pert}
 -\frac{q}{R^3} + \frac{3q\cos^2\theta}{2R^3} \equiv qW_1(R) +
 qW_2(R,\theta). 
 \end{equation}
Notice that the perturbation (\ref{pert}) corresponds to the case
$\beta=3$ considered in \cite{JMP}, but the unperturbed Hamiltonian
(\ref{H_0}) is indeed different. Without loss of generality, let us
assume hereafter that $\alpha = 1$. Notice that a dipole perturbation would give rise
to a Hamiltonian (\ref{ham}) with $W_1=0$ and $W_2=R^{-2}\sin\theta$, which indeed corresponds to a particular case of 
 the integrable case discussed in the Section 48 of \cite{dipole}.

In order to compute the Melnikov integrals \cite{Melnikov} for the Kepler problem with
quadrupole perturbations, we will adopt the 
 integral method adapted for parabolic orbits presented in 
\cite{{revisited}}. To this purpose, we need to obtain the
equivalent of the homoclinic orbit of our problem. The total energy
and the total angular momentum are the conserved quantities of our
system,  
\begin{subeqnarray}
\slabel{H_00}
  H_0 = \frac{\dot{R}^2}{2} + \frac{G^2}{2 R^2} - \frac{1}{R}, \\
  \slabel{G^2}
  G^2 = R^4 \dot{\theta}^2 + \frac{L_z ^2}{\cos^2 \theta},
\end{subeqnarray}
and from the expressions above, we have
\begin{subeqnarray}
  \frac{dR}{\sqrt{ 2 \left( H_0 + \frac{1}{R} \right) -
      \frac{G^2}{R^2} } } = \pm dt 
  \slabel{eqn-r}, \\
  \frac{1}{R^2} \frac{ dR }{ \sqrt{ 2 \left( H_0 +  \frac{1}{R} \right)
      - \frac{G^2}{R^2} } } = 
  \frac{ d \theta }{ \sqrt{ G^2 - \frac{L_z^2}{\cos^2 \theta} } }, 
  \slabel{eqn-theta}
\end{subeqnarray}
From (\ref{H_00}), we see that the minimum value of $R$ with $H_0 = 0$
satisfies $G^2 = 2 R_{min}$. We are interested in the parabolic orbits
and, hence, substituting $H_0 = 0$ in (\ref{eqn-r}) and
(\ref{eqn-theta}) and performing the integration, we have    
\begin{subeqnarray}
  \pm t = \frac{\sqrt{2}}{3} \left( R - \frac{R_{min}}{2} \right)
  \sqrt{R - R_{min}} + \rm{const}
  \slabel{t_de_r}, \\
  \frac{1}{4 A} \ln \left| \frac{A + \sin \theta}{A - \sin \theta}
  \right| + \rm{const} = \arctan \sqrt{\frac{R}{R_{min}} - 1}.
  \slabel{r_de_theta}
\end{subeqnarray}
where $A = \sqrt{ 1 - \frac{L_z ^2}{G^2} }$. Notice that, from
(\ref{G^2}), one has that $0 < A \le 1 $. 

Inverting
(\ref{t_de_r}), we have the expression $R (t)$ of the
homoclinoc orbit, but this is not necessary to our purposes. Also,
adjusting the constant in (\ref{r_de_theta}) so that $R = R_{min}$ for
$\theta = 0$, we have the following expression for $R (\theta)$ 
\begin{equation}
\label{R}
R (\theta) = R_{min} \sec^2 \left[ \frac{1}{4 A} \ln
  \left| \frac{A + \sin \theta}{A - \sin \theta} \right| \right]. 
\end{equation}
From  (\ref{R}), it is clear that $R(\theta)$ is an even function and
that the parabolic orbit can be parametrized with
$-\theta^*<\theta<\theta^*$, where $R(\theta^*)= \infty$, which leads
to  
\begin{equation}
\label{theta*}
\sin \theta^* = A \tanh A \pi. 
\end{equation}

The Melnikov conditions to detect integrability of a Hamiltonian
system of the type (\ref{H}) corresponds to the existence of simple
zeros for the quantities \cite{revisited}
\begin{equation}  
  M_1 (\theta_0) = \int_{- \infty}^{\infty} \{ H_0, W_2 \} \ud t,
  \qquad  
  M_2 (\theta_0) = \int_{- \infty}^{\infty} \{ G, W_2 \} \ud t,
\end{equation}
where the integrals are taken over the zero-energy manifold. For each
value of $A$, this is a two-dimensional manifold parametrized as
$\mathscr{R} = R (t - t_0)$ and $\vartheta = \Theta(t - t_0) +
\theta_0$, with arbitrary $t_0$ and $\theta_0$. We see that
\begin{equation} 
M_1 (\theta_0) = - \int_{- \infty}^{\infty}  \left. \left[
    \dot{\mathscr{R}} \frac{\partial W_2}{\partial R} + \dot{\Theta}
    \frac{\partial W_2}{\partial \theta} \right] \ud t = - W_2 \right|
_{t = - \infty}^{t = \infty} + \int_{- \infty}^{\infty} \frac{\partial
  W_2}{\partial t} \ud t = 0,
\end{equation}
and, with some trigonometry, that
\begin{equation} 
\label{M_2}
M_2 (\theta_0) =  - \frac{3}{8 R_{min}} \left[ I_1 \cos (2 \theta_0) +
  I_2 \sin (2 \theta_0) \right],
\end{equation}
where, after changing the integration variable,
\begin{equation} 
  I_1  = \int_{- \theta^*}^{\theta^*} \frac{\sin (2
    \theta)}{R(\theta)} \ud \theta, \qquad
  I_2 = \int_{- \theta^*}^{\theta^*} \frac{\cos (2
    \theta)}{R(\theta)} \ud \theta.
\end{equation}
Since $R (\theta)$ is an even function, we have $I_1 =
0$. Finally, the non identically zero contribution to the Melnikov
integral comes from the integral   
\begin{equation} \label{m_cond}
  I_2 = \frac{1}{R_{min}} \int_{-\theta^*}^{\theta^*} \cos
  (2\theta) \cos^2 \left[ \frac{1}{4A} \ln \left| \frac{A + \sin
        \theta}{A - \sin \theta} \right| \right] \ud \theta,
\end{equation}
where $\theta^*$ is given by  (\ref{theta*}).
Is is enough to prove that $I_2\ne 0$ for some value of $A$ to
establish that $M_2(\theta_0)$ given by (\ref{M_2}) has (infinitely
many) simple zeros, which implies the absence of the extra conserved
integral of motion and, consequently, that the motion is indeed
chaotic, irrespective of the sign of the perturbation parameter
$q$. Figure 1 depicts the integral (\ref{m_cond}) as a function of $A$,
and we can check that one has indeed $I_2\ne 0$ for $0<A<1$. 

\begin{figure}[h!]
\includegraphics[scale=0.5]{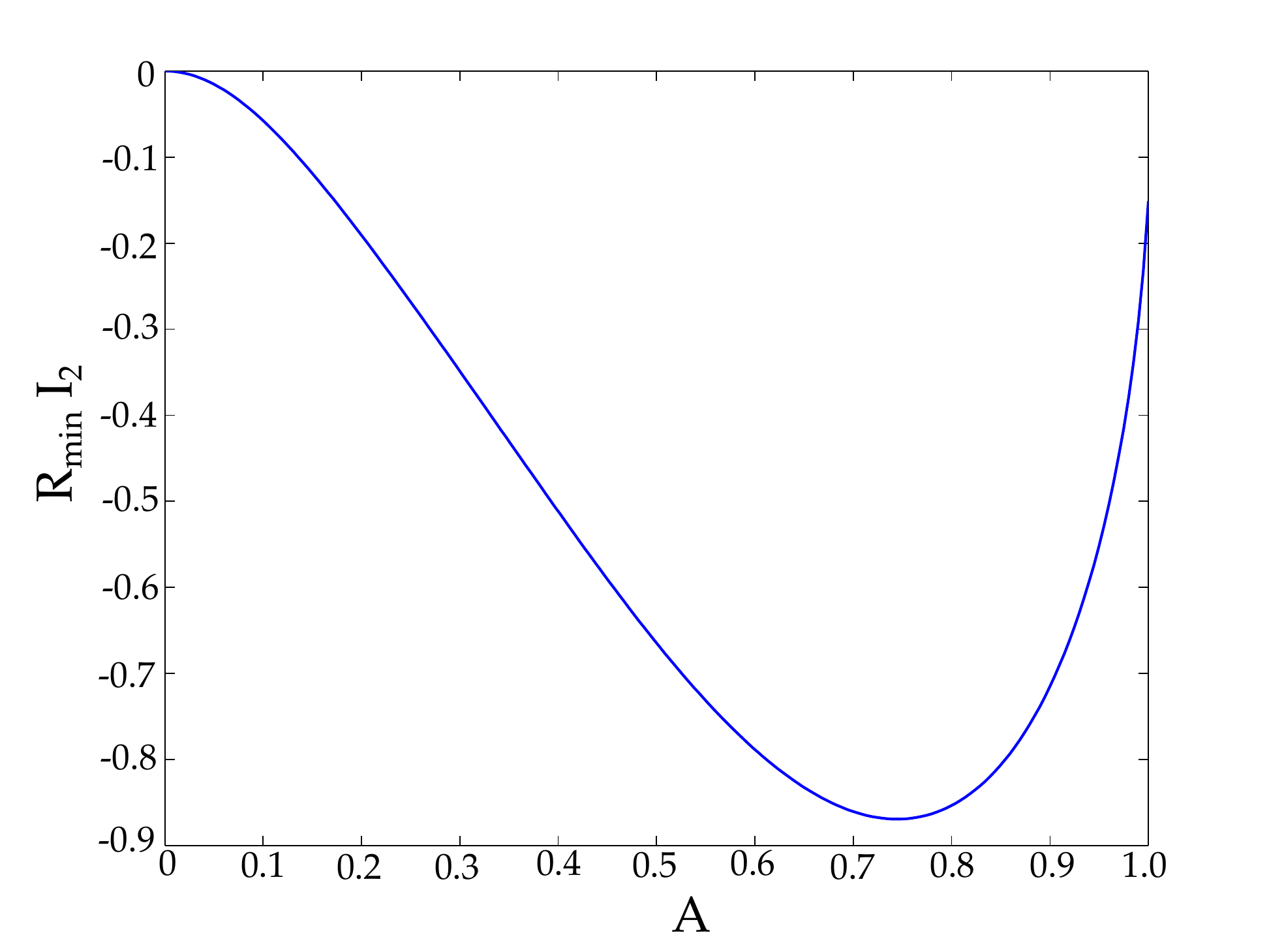}
\caption{The integral
  (\ref{m_cond})  for different values of
  $A$. Notice that $I_2 = 0$ only for $A = 0$ and for $A = 1$, the
  latter corresponding to $L_z=0$ and, thus, to the case $\beta=3$
  considered in \cite{JMP}.} 
\end{figure}

\section{Final remarks}

By using the Melnikov integral method adapted for parabolic orbits
\cite{{revisited}}, we prove   that the Hamiltonian flow on the
zero-energy manifold for the Kepler problem perturbed by a quadrupole
moment is chaotic, irrespective of the perturbation being of prolate
or oblate type. This result favors, in this way, the numerical results  obtained in 
\cite{LCF}, which are in conflict with those ones presented in 
 \cite{GL}.

\begin{acknowledgement}
The authors and grateful to FAPESP and CNPq for the financial support,
and  to the Fields Institute and the Universit\'e Libre de Bruxelles,
where part of this work was done, for the warm hospitality. GD would
like to thank M. Santoprete for the fruitful discussions at the Fields
Institute. 
\end{acknowledgement}

\end{document}